\documentclass[aps,pra,reprint,showpacs,floatfix,longbibliography]{revtex4-1}
\usepackage{graphicx, amsmath, amssymb}
\usepackage[caption=false]{subfig}

\begin{document}

%-------------------------------------------------------------------------------
% Shortcut Commands
%-------------------------------------------------------------------------------

\newcommand{\braket}[2]{{\left\langle #1 \middle| #2 \right\rangle}}
\newcommand{\bra}[1]{{\left\langle #1 \right|}}
\newcommand{\ket}[1]{{\left| #1 \right\rangle}}
\newcommand{\ketbra}[2]{{\left| #1 \middle\rangle \middle \langle #2 \right|}}

%-------------------------------------------------------------------------------
% Front Matter
%-------------------------------------------------------------------------------

\title{Irreconcilable Difference Between Quantum Walks and \\ Adiabatic Quantum Computing}

\author{Thomas G.~Wong}
	\affiliation{Faculty of Computing, University of Latvia, Rai\c{n}a bulv.~19, R\=\i ga, LV-1586, Latvia}
	\email{twong@lu.lv}

\author{David A.~Meyer}
	\affiliation{Department of Mathematics, University of California, San Diego, La Jolla, CA 92093-0112}
	\email{dmeyer@math.ucsd.edu}

\begin{abstract}
	Continuous-time quantum walks and adiabatic quantum evolution are two general techniques for quantum computing, both of which are described by Hamiltonians that govern their evolutions by Schr\"odinger's equation. In the former, the Hamiltonian is fixed, while in the latter, the Hamiltonian varies with time. As a result, their formulations of Grover's algorithm evolve differently through Hilbert space. We show that this difference is fundamental; they cannot be made to evolve along each other's path without introducing structure more powerful than the standard oracle for unstructured search. For an adiabatic quantum evolution to evolve like the quantum walk search algorithm, it must interpolate between three fixed Hamiltonians, one of which is complex and introduces structure that is stronger than the oracle for unstructured search. Conversely, for a quantum walk to evolve along the path of the adiabatic search algorithm, it must be a chiral quantum walk on a weighted, directed star graph with structure that is also stronger than the oracle for unstructured search. Thus the two techniques, although similar in being described by Hamiltonians that govern their evolution, compute by fundamentally irreconcilable means.
\end{abstract}


\maketitle

%-------------------------------------------------------------------------------
% Main Matter
%-------------------------------------------------------------------------------

\section{Introduction}

Grover's quantum search algorithm \cite{Grover1996} is ubiquitous in quantum information processing \cite{RP2011}, and its formulation into various quantum computing paradigms reveals major developments of the field. Grover's original unstructured search algorithm in 1996 was proposed in an era when the quantum circuit model was the only mainstream paradigm for quantum computing, so it consists of a series of quantum gates applied in discrete-time. When analog quantum computing was developed by Farhi and Gutmann in 1998 \cite{FG1998}, where the system evolves in continuous-time by Schr\"odinger's equation, Grover's algorithm was the first algorithm to be formulated in this new model. When quantum walks were popularized as algorithmic tools, the search problem was soon investigated \cite{SKW2003}, and Farhi and Gutmann's algorithm was shown to be a quantum walk \cite{CG2004}. Another model of analog quantum computing by Schr\"odinger evolution emerged in 2000 \cite{FGGS2000}, this time based on the adiabatic theorem with a time-dependent Hamiltonian. This was also developed by Farhi and Gutmann, in collaboration with Goldstone and Sipser, and was likely partly motivated by their previous work. Again, unstructured search was one of the first problems considered \cite{FGGS2000}, although it took some additional work by Roland and Cerf to get a square-root speedup \cite{RC2002}. Thus when new models for quantum computing develop, new formulations of Grover's algorithm immediately or soon follow.

Since Grover's algorithm is a common thread across each paradigm, it is natural to compare different formulations of it. Some prior work on such comparisons include Roland and Cerf's \cite{RC2003} connection between Grover's original circuit-based algorithm, Farhi and Gutmann's analog algorithm, and their local adiabatic algorithm; and Krovi, Ozols, and Roland's \cite{KOR2010} link between discrete-time quantum walks (or Markov chains) and adabatic quantum computing. In this paper, we focus on continuous-time quantum walks and adiabatic quantum computing, both of which evolve in continuous-time by Schr\"odinger's equation. But rather than focus on their similarities as in \cite{RC2003}, we focus their differences.

In continuous-time quantum walks, the Hamiltonian and its eigenstates are fixed throughout the evolution (with the exception of multi-stage quantum walks, which are fixed for each stage \cite{Wong7,Wong9,Wong16}, and nonlinear quantum walks, whose Hamiltonians vary to keep the eigenstates fixed \cite{Wong3,Wong4,Wong12}). In adiabatic quantum computing, however, the Hamiltonian is intentionally varied, such that the system stays in its instantaneous ground state \cite{FGGS2000}. Thus fixed eigenstates are used for one, and time-varying eigenstates are required for the other.

Comparing Grover's algorithm in these two models shows that they evolve through Hilbert space along significantly different paths. Here, we examine the reasons for this difference, showing why, under reasonable conditions, this difference cannot be overcome. To do this, we review Grover's algorithm \cite{Grover1996} in Section II, followed by Farhi and Gutmann's \cite{FG1998} quantum walk analogue in Section III, emphasizing that they follow far different paths \cite{Fenner2000}. This contrasts with Roland and Cerf's local adiabatic version \cite{RC2002} in Section IV, which does follow the same path as Grover's algorithm \cite{RC2003}. In Section V, we determine what adiabatic evolution follows the same path as Farhi and Gutmann's quantum walk search algorithm, showing that it necessitates a complex Hamiltonian that includes stucture that is far more powerful than the usual yes/no oracle. In Section VI, we consider the converse, showing that the quantum walk that follows the same path as Roland and Cerf's adiabatic search algorithm is Fenner's Hamiltonian construction \cite{Fenner2000}, which we interpret as a chiral quantum walk \cite{ZFKWLB2013,LBLLJBFZLBL2014} on the weighted, directed star graph and which also abandons the typical notion of an oracle. As part of the analysis, we improve upon Roland and Cerf's results \cite{RC2003} by showing that the ground state of their local adiabatic algorithm follows Grover's original path for all $N$, not just in the large $N$ limit. We conclude in Section VII, having shown that the two computational models are unable to simulate each other's unstructured search evolutions without abandoning the usual computational and oracular conditions.

%-------------------------------------------------------------------------------
% Section
%-------------------------------------------------------------------------------

\section{Grover's Original Algorithm}

\begin{figure}
\begin{center}
	\subfloat[]{
		\includegraphics[width=1.6in]{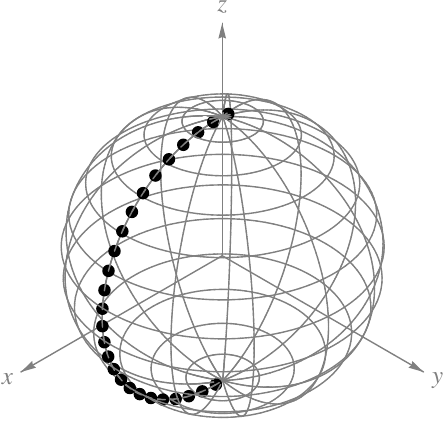}
		\label{fig:blochsphere_Grover}
	}
	\subfloat[]{
		\includegraphics[width=1.6in]{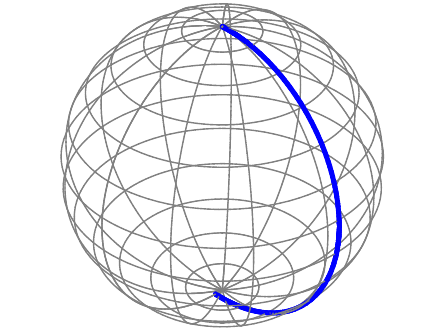}
		\label{fig:blochsphere_FG}
	}

	\subfloat[]{
		\includegraphics[width=1.6in]{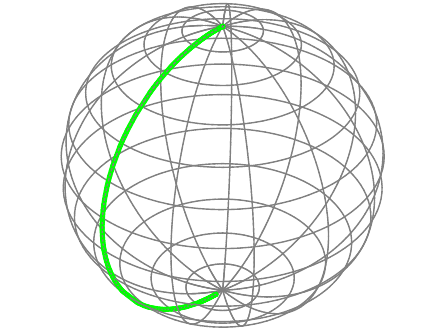}
		\label{fig:blochsphere_RC}
	}
	\subfloat[]{
		\includegraphics[width=1.6in]{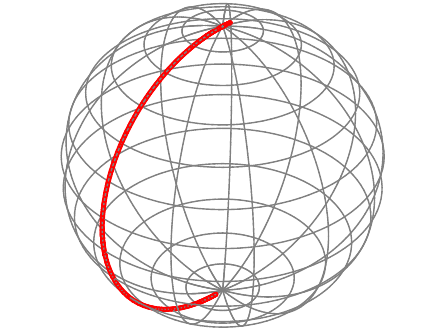}
		\label{fig:blochsphere_Fenner}
	}
	\caption{Evolution of quantum algorithms on the Bloch sphere with the marked vertex $\ket{w}$ at the North Pole, the equal superposition of unmarked vertices $\ket{r}$ at the South Pole, and $N = 1024$: (a) Grover's original discrete-time search algorithm, (b) Farhi and Gutmann's quantum walk analogue, (c) Roland and Cerf's adiabatic analogue, and (d) Fenner's quantum walk.}
\end{center}
\end{figure}

We begin by reviewing Grover's original discrete-time algorithm for solving the unstructured search problem. Given a computational basis $\{ \ket{1}, \dots, \ket{N} \}$ of an $N$-dimensional Hilbert space, the unstructured search problem is to find a ``marked'' basis state $\ket{w}$ by querying an oracle $R_w$ that flips the sign of $\ket{w}$ while leaving the other basis states alone. That is, $R_w$ is a reflection through $\ket{w}$. Note this oracle only responds ``yes/no'' as to whether a basis state is marked by applying a phase; it does not drive evolution between states alone. The system $\ket{\psi}$ begins in the equal superposition $\ket{s}$ of the basis states:
\[ \ket{s} = \frac{1}{\sqrt{N}} \sum_{i = 1}^N \ket{i}. \]
Grover's algorithm acts on this by repeatedly applying
\begin{equation}
	\label{eq:U}
	U = R_{s^\perp} R_w,
\end{equation}
where $R_{s^\perp}$ is a reflection through $\ket{s^\perp}$. These two reflections yield an overall rotation \cite{Aharonov1999,RP2011}, and the system evolves in a 2D subspace spanned by $\ket{w}$ and the equal superposition of unmarked states
\[ \ket{r} = \frac{1}{\sqrt{N-1}} \sum_{i \ne w} \ket{i}. \]
We can visualize the evolution in this 2D subspace as points on the Bloch sphere with $\ket{w}$ and $\ket{r}$ as the North and South Poles, respectively, as shown in Fig.~\ref{fig:blochsphere_Grover}. The system starts near the South Pole and takes fixed-length steps towards the North Pole. Since $\ket{\psi}$ is a superposition of $\ket{w}$ and $\ket{r}$ with real coefficients, the points lie on the Bloch sphere's $xz$-plane. After roughly $\pi\sqrt{N}/4$ applications of $U$, the system is rotated near the North Pole, so we have found $\ket{w}$ with probability near $1$.

%-------------------------------------------------------------------------------
% Section
%-------------------------------------------------------------------------------

\section{Search by Quantum Walk}

\begin{figure}
\begin{center}
	\includegraphics{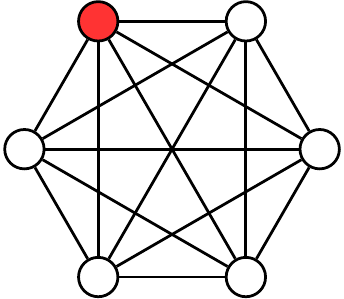}
	\caption{\label{fig:graph_complete} The complete graph with $N = 6$ vertices for Farhi and Gutmann's quantum walk search algorithm.}
\end{center}
\end{figure}

Now let us formulate Grover's algorithm in analog computational models, here as a quantum walk and next as an adiabatic evolution. As a quantum walk \cite{SKW2003}, the $N$ vertices of a graph can be used to label computational basis states $\{ \ket{1}, \dots, \ket{N} \}$, and we search for a marked vertex $\ket{w}$ by querying an oracle. In continuous-time \cite{CG2004}, the system $\ket{\psi}$ evolves from the equal superposition $\ket{s}$ over the vertices by Schr\"odinger's equation
\begin{equation}
	\label{eq:Schrodinger}
	i \frac{d}{dt} \ket{\psi} = H \ket{\psi},
\end{equation}
with Hamiltonian
\begin{equation}
	\label{eq:H_qwalk}
	H = -\gamma A - \ketbra{w}{w},
\end{equation}
where $\gamma$ is the jumping rate, $A$ is the adjacency matrix of the graph ($A_{ij} = 1$ if $i$ and $j$ are connected, and $0$ otherwise) that effects the quantum walk \cite{Wong19}, and $\ketbra{w}{w}$ is a Hamiltonian oracle \cite{Mochon2007} that marks the vertex to be found. This oracle is precisely the continuous-time version of the reflection $R_w$ in Grover's original discrete-time algorithm because alone it evolves the marked vertex by a phase, \textit{i.e.}, $e^{-i \ketbra{w}{w} t} \ket{w} = e^{-it} \ket{w}$, while leaving unmarked vertices unchanged. Thus it is the continuous-time version of a yes/no oracle.

Grover's problem is unstructured search, so it is search on the complete graph of $N$ vertices \cite{CG2004}, an example of which is shown in Fig.~\ref{fig:graph_complete}. In this case, the adjacency matrix has zeros on the diagonal and ones everywhere else. Since adding a multiple of the identity matrix only contributes a global, unobservable phase, we add $-\gamma I$ to the Hamiltonian, yielding $-\gamma(A+I) - \ketbra{w}{w}$, which is
\begin{equation}
	\label{eq:H_FG}
	H = - \gamma N \ketbra{s}{s} - \ketbra{w}{w}.
\end{equation}
When $\gamma = 1/N$ \cite{CG2004}, this Hamiltonian \eqref{eq:H_FG} equals the one introduced by Farhi and Gutmann to solve Grover's problem \footnote{This is actually the negative of Farhi and Gutmann's Hamiltonian, which contributes an unobservable, global phase, so we ignore it as did \cite{CG2004}.}, so their algorithm is actually a quantum walk on the complete graph. In this regard, Farhi and Gutmann's algorithm is the unique continuous-time quantum walk formulation of Grover's algorithm, up to a global phase. Using their results \cite{FG1998}, the Hamiltonian in the $\{ \ket{w}, \ket{r} \}$ basis is
\[ H = \frac{-1}{N} \begin{pmatrix}
	N+1 & \sqrt{N-1} \\
	\sqrt{N-1} & N-1 \\
\end{pmatrix}. \]
Applying the time-evolution operator $e^{-iHt}$ to the initial state $\ket{s}$, the state of the system at time $t$ is
\begin{equation}
	\label{eq:FG}
	\ket{\psi(t)} = e^{it} \begin{pmatrix}
		\frac{1}{\sqrt{N}} \cos\left(\frac{t}{\sqrt{N}}\right) + i \sin\left(\frac{t}{\sqrt{N}}\right) \\
		\sqrt{\frac{N-1}{N}} \cos\left(\frac{t}{\sqrt{N}}\right) \\
	\end{pmatrix}.
\end{equation}
Thus the system reaches a success probability of $1$ at time $\pi\sqrt{N}/2$.

Since the system evolves in the same 2D subspace spanned by $\{ \ket{w}, \ket{r} \}$, we again visualize the evolution on the Bloch sphere \cite{Wong6}, as shown in Fig.~\ref{fig:blochsphere_FG}. This reveals what was first pointed out by Fenner \cite{Fenner2000}: Farhi and Gutmann's algorithm evolves on a path far from, and slightly longer than, Grover's.

%-------------------------------------------------------------------------------
% Section
%-------------------------------------------------------------------------------

\section{Adiabatic Quantum Search}

Let us compare this to the adiabatic formulation of Grover's algorithm. In adiabatic quantum computing \cite{FGGS2000}, the system evolves by Schr\"odinger's equation \eqref{eq:Schrodinger} with time-dependent Hamiltonian
\begin{equation}
	\label{eq:H_adiabatic}
	H(s) = \left( 1 - s(t) \right) H_0 + s(t) H_f,
\end{equation}
where the interpolation schedule $s$ goes from $0$ to $1$ as the time $t$ goes from $0$ to the runtime $T$. This interpolates between the initial Hamiltonian $H_0$ and final Hamiltonian $H_f$. For the search problem, they are
\begin{equation}
	\label{eq:H0Hf}
	H_0 = I - \ketbra{s}{s}, \quad H_f = I - \ketbra{w}{w},
\end{equation}
so the initial ground state is the equal superposition $\ket{s}$, which is the initial state of the system, and the final ground state is the marked state $\ket{w}$ that we want to find. Note if we drop multiples of $I$ from this adiabatic Hamiltonian, divide it by $s$, and identify $\gamma N = (1-s)/s$, we get Farhi and Gutmann's quantum walk Hamiltonian \eqref{eq:H_FG}; this observation, however, does not yield a fundamental equivalence, since we now vary the Hamiltonian with time rather than keep it fixed. The system evolves in the same 2D subspace spanned by $\{ \ket{w}, \ket{r} \}$, in which the Hamiltonian \eqref{eq:H_adiabatic} is
\[ H(s) = \begin{pmatrix}
	(1-s) \frac{N-1}{N} & - (1-s) \frac{\sqrt{N-1}}{N} \\
	- (1-s) \frac{\sqrt{N-1}}{N} & 1 - (1-s) \frac{N-1}{N} \\
\end{pmatrix}. \]
The (unnormalized) eigenvectors of this are
\begin{equation}
	\label{eq:adiabatic_eigenstates}
	\psi_{0,1}(s) = \frac{2(1-s) - N(1-2s) \pm N g}{2\sqrt{N-1}(1-s)} \ket{w} + \ket{r}
\end{equation}
with energy gap
\begin{equation}
	\label{eq:gap}
	g(s) = \sqrt{\frac{N - 4(N-1)s(1-s)}{N}}.
\end{equation}

\begin{figure}
\begin{center}
	\subfloat[]{
		\includegraphics{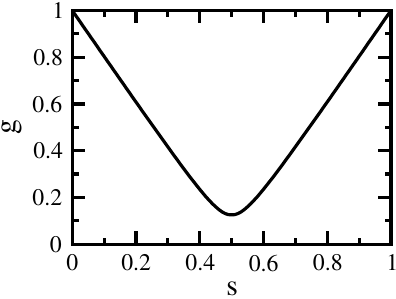}
		\label{fig:adiabatic_gap}
	} \enspace
	\subfloat[]{
		\includegraphics{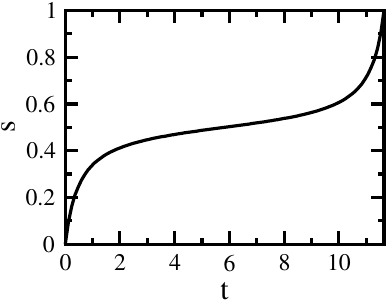}
		\label{fig:adiabatic_schedule} 
	}
	\caption{For adiabatic quantum search with $N = 64$: (a) the energy gap $g(s)$ in \eqref{eq:gap} for $s \in [0,1]$, and (b) Roland and Cerf's schedule $s(t)$ as $t \in [0:T]$ with $\epsilon = 1$.}
\end{center}
\end{figure}

The adiabatic theorem says that if the system evolves ``much slower'' than the reciprocal of the energy gap squared \cite{Messiah1999}, then the system stays in its instantaneous ground state throughout its evolution. For search, this means the system would evolve from $\ket{s}$ to $\ket{w}$, succeeding. The energy gap \eqref{eq:gap} is shown in Fig.~\ref{fig:adiabatic_gap}, and Roland and Cerf showed that the interpolation schedule
\begin{align}
	t = \frac{N}{2\epsilon\sqrt{N-1}} \Big\{ &\tan^{-1} \left[\sqrt{N-1}(2s-1)\right] \notag \\
	&+ \tan^{-1} \left( \sqrt{N-1} \, \right) \Big\}, \label{eq:schedule}
\end{align}
whose inverse is plotted in Fig.~\ref{fig:adiabatic_schedule}, locally satisfies the adiabatic theorem, so it evolves quickly when the gap is large and slowly when the gap is small \cite{RC2002}. With this schedule, the algorithm takes total time
\begin{equation}
	\label{eq:adiabatic_runtime}
	T = \frac{N}{\epsilon\sqrt{N-1}} \tan^{-1} \left(\sqrt{N-1}\right) \approx \frac{\pi}{2\epsilon} \sqrt{N}
\end{equation}
to evolve to its final state $\ket{\psi(s=1)}$. From Rezakhani, Pimachev, and Lidar's \cite{RPL2010} explicit evolution of the time-dependent Schr\"odinger equation, when the runtime scales as
\[ T = O \left( \sqrt{N} \ln \left( \frac{1}{\delta} \right) \right), \]
then the adiabatic error, which measures how far the system is from its true final ground state $\ket{w}$, is upper bounded by $\delta$:
\begin{equation}
	\label{eq:adiabatic_error}
	\sqrt{1 - \left| \braket{w}{\psi(s=1)} \right|^2} \le \delta.
\end{equation}
Thus evolving with constant $\epsilon$ yields an $O(\sqrt{N})$ runtime, which implies that $\ln(1/\delta)$ is a nonzero constant, so $\delta$ is a constant less than $1$. Thus the adiabatic error is upper bounded by a constant less than $1$, and we have an $O(\sqrt{N})$ search algorithm, even with an expected constant number of repetitions, on average.

Although constant $\epsilon$ yields an $O(\sqrt{N})$ search algorithm, the state may be far from its instantaneous ground state $\psi_0(s)$ from \eqref{eq:adiabatic_eigenstates}. If $\epsilon$ scales less than a constant so that it decreases with $N$, however, then it does follow $\psi_0(s)$ for large $N$. Since $\psi_0(s)$ is spanned by $\{ \ket{w}, \ket{r} \}$, we plot its evolution on the Bloch sphere in Fig.~\ref{fig:blochsphere_RC}, and it stays in the $xz$-plane because it always has real coefficients, following the same path as Grover's algorithm \cite{RC2003}.

%-------------------------------------------------------------------------------
% Section
%-------------------------------------------------------------------------------

\section{Adiabatic Evolution Following Quantum Walk Search}

We have seen that the adiabatic algorithm follows the same path as Grover's original algorithm, but is far from Farhi and Gutmann's quantum walk. Now we determine what adiabatic algorithm does follow the same path as the quantum walk search algorithm in Fig.~\ref{fig:blochsphere_FG}. To do this, we find what time-dependent Hamiltonian $H(t)$ has \eqref{eq:FG} as its ground state.

To simplify the notation, we drop the global phase of \eqref{eq:FG} and name its amplitudes $\alpha$ and $\beta$, \textit{i.e.},
\begin{gather*}
	\alpha(t) = \frac{1}{\sqrt{N}} \cos\left(\frac{t}{\sqrt{N}}\right) + i \sin\left(\frac{t}{\sqrt{N}}\right) \\
	\beta(t) = \sqrt{\frac{N-1}{N}} \cos\left(\frac{t}{\sqrt{N}}\right).
\end{gather*}
Then we want the ground state of the adiabatic Hamiltonian $H(t)$ to be
\[ \ket{\psi_0(t)} = \alpha(t) \ket{w} + \beta(t) \ket{r}. \]
The other eigenstate of $H(t)$, which is its first and only excited state, is orthogonal to this, so it is
\[ \ket{\psi_1(t)} = \beta(t) \ket{w} - \alpha^*(t) \ket{r}. \]
Say these eigenstates have respective eigenvalues $\lambda_0$ and $\lambda_1$. Then since the eigenvectors diagonalize the Hamiltonian,
\begin{align*}
	H(t) 
	&= \lambda_0 \ketbra{\psi_0}{\psi_0} + \lambda_1 \ketbra{\psi_1}{\psi_1} \\
	&= \lambda_0 \begin{pmatrix}
		|\alpha|^2 & \alpha\beta \\
		\alpha^*\beta & \beta^2 \\
	   \end{pmatrix} + \lambda_1 \begin{pmatrix}
		\beta^2 & -\alpha^*\beta \\
		-\alpha\beta & |\alpha|^2 \\
	   \end{pmatrix} \\
	&= \begin{pmatrix}
		\lambda_0 |\alpha|^2 + \lambda_1 \beta^2 & (\lambda_0 - \lambda_1)\alpha\beta \\
		(\lambda_0 - \lambda_1)\alpha^*\beta & \lambda_0 \beta^2 + \lambda_1 |\alpha|^2
	   \end{pmatrix}. 
\end{align*}
Usually, adiabatic algorithms have real Hamiltonians, such as the unstructured search algorithm \eqref{eq:H_adiabatic} and \eqref{eq:H0Hf} in the last section. But here, the only way to make the Hamiltonian real is for $\lambda_0 = \lambda_1$, but this detrimentally eliminates the energy gap and causes the Hamiltonian to be the identity matrix. So the adiabatic Hamiltonian must be complex. For simplicity, we choose $\lambda_0 = -\lambda_1$, which is always possible because we can add a multiple of the identity matrix to the Hamiltonian to redefine the zero of energy. Then the Hamiltonian is
\[ H(t) = \lambda_1 \begin{pmatrix}
	-|\alpha|^2 + \beta^2 & -2\alpha\beta \\
	-2\alpha^*\beta & |\alpha|^2 - \beta^2 \\
\end{pmatrix}. \]
Plugging in for $\alpha$ and $\beta$, the Hamiltonian reduces to a pleasant interpolation between three terms:
\[ H(s) = \lambda_1(s) \left[ (1 - s) H_0 + s H_f + \sqrt{s(1-s)} H_e \right], \]
where
\[ s(t) = \sin^2\left(\frac{t}{\sqrt{N}}\right) \]
is the interpolation schedule and
\begin{gather*}
	H_0 = \begin{pmatrix}
		\frac{N-2}{N} & -\frac{2\sqrt{N-1}}{N} \\
		-\frac{2\sqrt{N-1}}{N} & -\frac{N-2}{N} \\
	\end{pmatrix}, \quad
	H_f = \begin{pmatrix}
		-1 & 0 \\
		0 & 1 \\
	\end{pmatrix}, \\
	H_e = \begin{pmatrix}
		0 & -2i\sqrt{\frac{N-1}{N}} \\
		2i\sqrt{\frac{N-1}{N}} & 0 \\
	\end{pmatrix}
\end{gather*}
are respectively the beginning, final, and ``extra'' Hamiltonians. Having a third, extra Hamiltonian is a common technique for manipulating the evolution path of adiabatic algorithms to avoid a small gap in the space of Hamiltonians \cite{FGG2002,HS2014,Crosson2014}, although it typically carries a factor of $s(1-s)$, whereas ours contains a square-root.

Before commenting on the consequences of this evolution, we still need to find $\lambda_1$ in terms of $s$. Differentiating the interpolation schedule $s(t)$ and using the adiabatic theorem $ds/dt = \epsilon g^2(s)$,
\[ 2 \sin\left(\frac{t}{\sqrt{N}}\right) \cos\left(\frac{t}{\sqrt{N}}\right) \frac{1}{\sqrt{N}} = \epsilon g^2(s). \]
So the energy gap is
\[ g = \sqrt{\frac{2 \sqrt{s(1-s)}}{\epsilon \sqrt{N}}} \]
Since $\lambda_0 = -\lambda_1$, the energy gap is $g(s) = 2 \lambda_1$, or $\lambda_1 = g(s) / 2$. Thus
\[ \lambda_1(s) = \sqrt[4]{ \frac{s(1-s)}{4\epsilon^2N} }. \]
Putting everything together, the adiabatic Hamiltonian that follows the quantum walk search algorithm's evolution is
\[ H(s) = \sqrt[4]{ \frac{s(1-s)}{4\epsilon^2N} } \left[ (1 - s) H_0 + s H_f + \sqrt{s(1-s)} H_e \right]. \]

Now let us discuss the consequences of this Hamiltonian, aside from it being complex rather than real. Consider each of the three Hamiltonians comprising it. $H_0$ has ground state $\ket{s}$ and excited state $\ket{s^\perp}$ with respective eigenvalues $-1$ and $1$. Thus it can be written as
\[ H_0 = \ketbra{s^\perp}{s^\perp} - \ketbra{s}{s}. \]
Similarly, $H_f$ has ground state $\ket{w}$ and excited state $\ket{r}$ with respective eigenvalues $-1$ and $1$, so it is
\[ H_f = \ketbra{r}{r} - \ketbra{w}{w}. \]
These look fairly similar to the initial and final Hamiltonians of the standard adiabatic quantum search algorithm \eqref{eq:H0Hf}, which had the same eigenvectors, but with eigenvalues $0$ and $1$. These similarities are deceiving, however. The interpolation schedule here is very different from the unstructured search algorithm's \eqref{eq:schedule}, and most strikingly different is the extra Hamiltonian, which can be written as
\[ H_e = 2i\sqrt{\frac{N-1}{N}} \left( \ketbra{r}{w} - \ketbra{w}{r} \right). \]
This extra term changes everything. Rather than just having the $\ketbra{w}{w}$ term in $H_f$ serving as a standard yes/no Hamiltonian oracle \cite{Mochon2007}, $H_e$ acts much more powerfully. It introduces structure that drives evolution between $\ket{w}$ and $\ket{r}$, rather than just applying a phase to $\ket{w}$. The power of this term is evident in the operator norm of $H(s)$, which is $\Theta(1/N^{1/4})$ compared to $\Theta(1)$ for the unstructured search Hamiltonian \eqref{eq:H_adiabatic} and \eqref{eq:H0Hf}. So if $H(s)$ were rescaled to have constant operator norm, it would have a constant energy gap and find $\ket{w}$ in constant time. This confirms that the ``oracle'' is no longer a standard yes/no one, for which Grover's $\Theta(\sqrt{N})$ runtime is optimal \cite{BBBV1997,FG1998}. Thus while there exists an adiabatic evolution that follows the same path as the quantum walk search algorithm, it abandons the typical notion of an oracle and so does not solve the search problem itself.

Since Farhi and Gutmann's algorithm is, up to a global phase, the unique continuous-time quantum walk formulation of Grover's algorithm, this suffices to prove that adiabatic quantum computing is unable to solve Grover's algorithm in the same way as the quantum walk. Thus the two models compute by fundamentally irreconcilable means.

This irreconcilability is not a judgment on their computational power, of course. Both models are universal for quantum computing and are polynomially equivalent to the standard gate model \cite{Aharonov2004,Childs2013}. Both solve Grover's problem in $O(\sqrt{N})$ time. Yet our result illustrates that \emph{how} they compute is different, even though \emph{what} they compute is the same.

%-------------------------------------------------------------------------------
% Section
%-------------------------------------------------------------------------------

\section{Quantum Walk Following Adiabatic Quantum Search}

For completeness, we now consider the converse: what quantum walk follows the same evolution as the adiabatic quantum search algorithm in Fig.~\ref{fig:blochsphere_RC}? Of course, we already know that the resulting quantum walk will not solve Grover's problem, but it is enlightening to see how it deviates from Farhi and Gutmann's algorithm.

In response to the observation that Farhi and Gutmann's quantum walk search algorithm in Fig.~\ref{fig:blochsphere_FG} evolves far from Grover's in Fig.~\ref{fig:blochsphere_Grover}, Fenner \cite{Fenner2000} gave an alternative Hamiltonian
\begin{equation}
	\label{eq:H_Fenner}
	H_\text{F} = \frac{i}{\sqrt{N}} ( \ketbra{w}{s} - \ketbra{s}{w} )
\end{equation}
that for some time interval exactly applies Grover's iterate \eqref{eq:U} \footnote{Fenner's Hamiltonian in \cite{Fenner2000} is actually twice this; ours is closer in form and runtime to Farhi and Gutmann's.}. This also evolves in the subspace spanned by $\{ \ket{w}, \ket{r} \}$, and visualizing the evolution governed by this Hamiltonian on the Bloch sphere in Fig.~\ref{fig:blochsphere_Fenner} shows that it follows the same path as Grover's algorithm, as expected.

Let us explicitly find the state $\ket{\psi_\text{F}(t)}$ as it evolves by Fenner's Hamiltonian \eqref{eq:H_Fenner}. In the $\{ \ket{w}, \ket{r} \}$ basis,
\[ H_\text{F} = \frac{i}{N} \begin{pmatrix}
	0 & \sqrt{N-1} \\
	-\sqrt{N-1} & 0 \\
\end{pmatrix}. \]
Then the time-evolution operator
\begin{equation}
	\label{eq:Fenner_rotation}
	e^{-iH_\text{F}t} = \begin{pmatrix}
		\cos\left(\frac{\sqrt{N-1}}{N}t\right) & \sin\left(\frac{\sqrt{N-1}}{N}t\right) \\
		-\sin\left(\frac{\sqrt{N-1}}{N}t\right) & \cos\left(\frac{\sqrt{N-1}}{N}t\right) \\
	\end{pmatrix}
\end{equation}
is simply a rotation by $\sqrt{N-1}t/N$. Applying this to the initial state $\ket{s}$, the state of the system at time $t$ is
\[
	\ket{\psi_\text{F}(t)} = \begin{pmatrix}
			\frac{1}{\sqrt{N}} \cos \left( \frac{\sqrt{N-1}}{N} t \right) + \frac{\sqrt{N-1}}{\sqrt{N}} \sin \left( \frac{\sqrt{N-1}}{N} t \right) \\
			\frac{\sqrt{N-1}}{\sqrt{N}} \cos \left( \frac{\sqrt{N-1}}{N} t \right) - \frac{1}{\sqrt{N}} \sin \left( \frac{\sqrt{N-1}}{N} t \right) \\
	\end{pmatrix}.
\]
So the system reaches success probability $1$ at time roughly $\pi\sqrt{N}/2$. 
Since this state has real amplitudes, it stays in the $xz$-plane of the Bloch sphere. For important use later, we unnormalize the state so that the coefficient of $\ket{r}$ is $1$:
\begin{equation}
	\label{eq:state_Fenner}
	\psi_\text{F}(t) = \frac{\cos \left( \frac{\sqrt{N-1}}{N} t \right) + \sqrt{N-1} \sin \left( \frac{\sqrt{N-1}}{N} t \right)}{\sqrt{N-1} \cos \left( \frac{\sqrt{N-1}}{N} t \right) - \sin \left( \frac{\sqrt{N-1}}{N} t \right)} \ket{w} + \ket{r}.
\end{equation}

\begin{figure}
\begin{center}
	\includegraphics{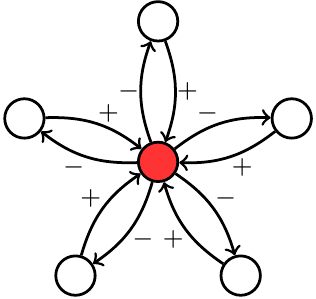}
	\caption{\label{fig:graph_Fenner} The weighted, directed star graph with $N = 6$ vertices for Fenner's Hamiltonian, where the $\pm$ weights indicate $\pm i/N$.}
\end{center}
\end{figure}

While Fenner's Hamiltonian \eqref{eq:H_Fenner} does not take the form of a typical quantum walk \eqref{eq:H_qwalk}, we show that it still has local transitions and is a type of quantum walk. The Hamiltonian \eqref{eq:H_Fenner} acts on computational basis states by
\[ H_\text{F} \ket{i} = \begin{cases}
	\frac{-i}{N} \sum_{j \ne w} \ket{j}, & i = w \\
	\frac{i}{N} \ket{w}, & i \ne w \\
\end{cases}. \]
Thus it takes amplitude from the marked vertex and transitions it to the other vertices with a factor of $-i/N$, and takes amplitude from the non-marked vertices and transitions it to the marked vertex with a factor of $i/N$. This can be drawn as shown in Fig.~\ref{fig:graph_Fenner}. Thus Fenner's Hamiltonian effects a quantum walk on the star graph, with the central node ``marked,'' and with the edges directed and weighted so that leaving the central node has a weight of $-i/N$ and going into the central node has weight of $i/N$. Since these directions have conjugate phases $\pm i = e^{\pm i\pi/2}$, it is a chiral quantum walk, which breaks time-reversal symmetry \cite{ZFKWLB2013,LBLLJBFZLBL2014}. Typical quantum walk search Hamiltonians have the quantum walk and oracle as separate terms as in \eqref{eq:H_qwalk}, but with Fenner's they are intertwined \eqref{eq:H_Fenner}. Essentially, the ``oracle'' comes from the structure of the graph, where the search problem is to find the vertex with preferential treatment, \textit{i.e.}, the central node of the star graph. This is not a yes/no oracle as with regular search problems. This makes the ``oracle'' much more powerful. In fact, the operator norm of Fenner's Hamiltonian is $1/\sqrt{N}$, in contrast to Farhi and Gutmann's $1$. So if we rescaled Fenner's Hamiltonian to be norm 1, it would search in constant time, which is unsurprising because a classical random walk would jump from an arm of the star graph in Fig.~\ref{fig:graph_Fenner} to the marked center in one step, and the optimality of Grover's algorithm is for yes/no oracles \cite{BBBV1997,FG1998}, which Fenner's is not. Even with these differences, it is still a quantum walk, albeit an atypical one that does not solve unstructured search. Note it is possible to define a chiral quantum walk search algorithm for unstructured search that retains the usual yes/no oracle term \cite{Wong14}, but that differs from the structure here.

Furthermore, any quantum walk algorithm that evolves with real amplitudes, staying in the $xz$-plane of the Bloch sphere as adiabatic algorithms typically do, requires that the Hamiltonian be purely imaginary so that the time-evolution operator $e^{-iHt}$ is real. But since the Hamiltonian must also be Hermitian, this implies that the terms on the diagonal must be real, not imaginary, which means they must be zero. This excludes the standard oracle $\ketbra{w}{w}$ in \eqref{eq:H_FG} because it would be on the diagonal. In addition, the off-diagonal terms must have conjugate phases $\pm i$ as in Fenner's \eqref{eq:H_Fenner} so that $H$ is Hermitian. Thus for a quantum walk to evolve with real amplitudes like the adiabatic quantum search algorithm, it must be a chiral quantum walk and abandon the usual oracle. This reveals how the quantum walk must deviate from Farhi and Gutmann's algorithm in order to simulate the adiabatic search algorithm.

We have shown that the ground state of Roland and Cerf's adiabatic evolution and the state of Fenner's quantum walk both follow the same path. Now we give two different arguments showing that their speeds are also identical when using Roland and Cerf's schedule \eqref{eq:schedule} with $\epsilon = 1$.

The first argument comes from analyzing the rotation rates. As shown by Roland and Cerf in Eqs.~(32) and (37) of \cite{RC2003}, their adiabatic algorithm with schedule \eqref{eq:schedule} rotates with constant angular velocity $\epsilon \sqrt{N-1} / N$ for large $N$. From \eqref{eq:Fenner_rotation}, Fenner's algorithm applies rotations at constant angular velocity $\sqrt{N-1} / N$. Thus they yield the same rotation rate when $\epsilon = 1$ for large $N$.

The second argument proves an exact algebraic equivalence, which is true for all $N$, not just when it is large. Beginning with Fenner's (unnormalized) state in \eqref{eq:state_Fenner}, we multiply the top and bottom of the $\ket{w}$ coefficient by $\sin(\cdot)$, where the dot indicates $\sqrt{N-1}t/N$:
\begin{equation}
	\label{eq:Fenner_scaled}
	\psi_\text{F}(t) = \frac{\sin(\cdot) \cos(\cdot) + \sqrt{N-1} \sin^2(\cdot)}{\sqrt{N-1} \sin(\cdot) \cos(\cdot) - \sin^2(\cdot)} \ket{w} + \ket{r}.
\end{equation}
Now we want to substitute Roland and Cerf's interpolation schedule \eqref{eq:schedule} with $\epsilon = 1$, showing that it equals the adiabatic algorithm's ground state. To do this, we rewrite the schedule using the trigonometric identity
\[ \tan^{-1} (\alpha) + \tan^{-1} (\beta) = \tan^{-1} \left( \frac{\alpha + \beta}{1 - \alpha\beta} \right) \mod \pi, \]
which causes it to become \cite{RC2003}
\[ t = \begin{cases}
	\frac{1}{2} \frac{N}{\sqrt{N-1}} \; \tan^{-1} \left( \frac{2s\sqrt{N-1}}{1-(N-1)(2s-1)} \right), & s \le \frac{1}{2} \\
	\frac{1}{2} \frac{N}{\sqrt{N-1}} \! \left[ \tan^{-1} \left( \frac{2s\sqrt{N-1}}{1-(N-1)(2s-1)} \right) + \pi \right], & s > \frac{1}{2} \\
\end{cases}. \]
Let us call the arctangent factor in this $\theta$:
\[ \theta = \tan^{-1} \left( \frac{2s\sqrt{N-1}}{1-(N-1)(2s-1)} \right) + (\pi), \]
where the $\pi$ is added when $s > 1/2$. Then
\[ \cos\theta = \frac{2s - N(2s-1)}{\sqrt{N}\sqrt{N(2s-1)^2 + 4s(1-s)}} = \frac{2s - N(2s-1)}{Ng}. \]
Using this, $\cos(\cdot)$ and $\sin(\cdot)$ in \eqref{eq:Fenner_scaled} are:
\begin{gather*}
	\cos(\cdot) = \cos \frac{\theta}{2} = \sqrt{\frac{1+\cos\theta}{2}} = \sqrt{\frac{Ng + 2s - N(2s-1)}{2Ng}}, \\
	\sin(\cdot) = \sin \frac{\theta}{2} = \sqrt{\frac{1-\cos\theta}{2}} = \sqrt{\frac{Ng - 2s + N(2s-1)}{2Ng}} .
\end{gather*}
Then
\begin{gather*}
	\sin(\cdot) \cos(\cdot) = \frac{2s\sqrt{N-1}}{2Ng}, \\
	\sin^2(\cdot) = \frac{Ng - 2s + N(2s-1)}{2Ng}.
\end{gather*}
Plugging these into Fenner's state \eqref{eq:Fenner_scaled},
\begin{align*}
	\psi_\text{F}(s) &= \frac{2s\sqrt{N-1} + \sqrt{N-1} \left[ Ng - 2s + N(2s-1) \right]}{2s(N-1) - \left[ Ng - 2s + N(2s-1) \right]} \ket{w} \\ &\quad+ \ket{r}.
\end{align*}
Simplifying and rearranging,
\begin{align*}
	\psi_\text{F}(s) 
	&= -\sqrt{N-1} \frac{g-1+2s}{g-1} \ket{w} + \ket{r}\\
	&= -\sqrt{N-1} \left( 1 + \frac{2s(g+1)}{g^2-1} \right) \ket{w} + \ket{r}.
\end{align*}
Plugging in for $g^2$ using \eqref{eq:gap} and simplifying,
\begin{align*}
	\psi_\text{F}(s) 
	&= -\sqrt{N-1} \frac{2s \left[ 2-2s+N(-1+g+2s) \right]}{-4(N-1)(1-s)s} \ket{w} + \ket{r} \\
	&= \frac{2(1-s) - N(1-2s) + Ng}{2\sqrt{N-1}(1-s)} \ket{w} + \ket{r}.
\end{align*}
But from \eqref{eq:adiabatic_eigenstates}, this is exactly $\psi_0(s)$, the ground state of the adiabatic algorithm. Thus the ground state of Roland and Cerf's local adiabatic algorithm with $\epsilon = 1$ evolves identically to Fenner's quantum walk. Put another way, equating $\psi_0(s)$ \eqref{eq:adiabatic_eigenstates} and $\psi_\text{F}(t)$ \eqref{eq:state_Fenner}, then solving for $t$ is terms of $s$, yields the schedule \eqref{eq:schedule} with $\epsilon = 1$.

As previously discussed, since the adiabatic algorithm follows its ground state up to adiabatic error, $\epsilon$ must scale less than a constant for it to closely follow its ground state. Choosing this slows down the adiabatic evolution, but its state (up to adiabatic error) equals Fenner's quantum walk when both algorithms are the same fraction of the way through their evolutions. If desired, Fenner's Hamiltonian can also be rescaled so that the two algorithms evolve at the same speed.

%-------------------------------------------------------------------------------
% Section
%-------------------------------------------------------------------------------

\vspace{0.25in} % Fix lack of vertical space.
\section{Conclusion}

We have shown that the quantum walk and adiabatic quantum computing formulations of Grover's algorithm solve the unstructured search problem through fundamentally irreconcilable means. For an adiabatic evolution to follow the same path as Farhi and Gutmann's quantum walk search algorithm, which is the unique continuous-time quantum walk formulation of Grover's algorithm, the Hamiltonian must be complex, not real, and introduce structure that is beyond the standard yes/no oracle that search problems presume. Similarly, for a quantum walk to follow the same path as Roland and Cerf's adiabatic quantum search algorithm, it must be a chiral quantum walk with zeros on the diagonal that also introduces structure more powerful than the usual oracle. Thus the two quantum computational models can only simulate each other's unstructured search algorithms by abandoning the usual computational and oracular conditions.

%-------------------------------------------------------------------------------
% Acknowledgments
%-------------------------------------------------------------------------------

\begin{acknowledgments}
	Our great appreciation goes to Eleanor Rieffel for her valuable feedback and advice in developing this work.
	T.W.~was supported by the European Union Seventh Framework Programme (FP7/2007-2013) under the QALGO (Grant Agreement No.~600700) project, and the ERC Advanced Grant MQC.
	D.M.~was partially supported by the Air Force Office of Scientific Research as part of the Transformational Computing in Aerospace Science and Engineering Initiative under grant FA9550-12-1-0046.
\end{acknowledgments}

%-------------------------------------------------------------------------------
% References.
%-------------------------------------------------------------------------------

\bibliography{refs}

%merlin.mbs apsrev4-1.bst 2010-07-25 4.21a (PWD, AO, DPC) hacked
%Control: key (0)
%Control: author (0) dotless jnrlst
%Control: editor formatted (1) identically to author
%Control: production of article title (0) allowed
%Control: page (1) range
%Control: year (0) verbatim
%Control: production of eprint (0) enabled
\begin{thebibliography}{33}%
\makeatletter
\providecommand \@ifxundefined [1]{%
 \@ifx{#1\undefined}
}%
\providecommand \@ifnum [1]{%
 \ifnum #1\expandafter \@firstoftwo
 \else \expandafter \@secondoftwo
 \fi
}%
\providecommand \@ifx [1]{%
 \ifx #1\expandafter \@firstoftwo
 \else \expandafter \@secondoftwo
 \fi
}%
\providecommand \natexlab [1]{#1}%
\providecommand \enquote  [1]{``#1''}%
\providecommand \bibnamefont  [1]{#1}%
\providecommand \bibfnamefont [1]{#1}%
\providecommand \citenamefont [1]{#1}%
\providecommand \href@noop [0]{\@secondoftwo}%
\providecommand \href [0]{\begingroup \@sanitize@url \@href}%
\providecommand \@href[1]{\@@startlink{#1}\@@href}%
\providecommand \@@href[1]{\endgroup#1\@@endlink}%
\providecommand \@sanitize@url [0]{\catcode `\\12\catcode `\$12\catcode
  `\&12\catcode `\#12\catcode `\^12\catcode `\_12\catcode `\%12\relax}%
\providecommand \@@startlink[1]{}%
\providecommand \@@endlink[0]{}%
\providecommand \url  [0]{\begingroup\@sanitize@url \@url }%
\providecommand \@url [1]{\endgroup\@href {#1}{\urlprefix }}%
\providecommand \urlprefix  [0]{URL }%
\providecommand \Eprint [0]{\href }%
\providecommand \doibase [0]{http://dx.doi.org/}%
\providecommand \selectlanguage [0]{\@gobble}%
\providecommand \bibinfo  [0]{\@secondoftwo}%
\providecommand \bibfield  [0]{\@secondoftwo}%
\providecommand \translation [1]{[#1]}%
\providecommand \BibitemOpen [0]{}%
\providecommand \bibitemStop [0]{}%
\providecommand \bibitemNoStop [0]{.\EOS\space}%
\providecommand \EOS [0]{\spacefactor3000\relax}%
\providecommand \BibitemShut  [1]{\csname bibitem#1\endcsname}%
\let\auto@bib@innerbib\@empty
%</preamble>
\bibitem [{\citenamefont {Grover}(1996)}]{Grover1996}%
  \BibitemOpen
  \bibfield  {author} {\bibinfo {author} {\bibfnamefont {L.~K.}\ \bibnamefont
  {Grover}},\ }\bibfield  {title} {\enquote {\bibinfo {title} {A fast quantum
  mechanical algorithm for database search},}\ }in\ \href@noop {} {\emph
  {\bibinfo {booktitle} {Proceedings of the 28th Annual ACM Symposium on Theory
  of Computing}}},\ \bibinfo {series and number} {STOC '96}\ (\bibinfo
  {publisher} {ACM},\ \bibinfo {address} {New York, NY, USA},\ \bibinfo {year}
  {1996})\ pp.\ \bibinfo {pages} {212--219}\BibitemShut {NoStop}%
\bibitem [{\citenamefont {Rieffel}\ and\ \citenamefont {Polak}(2011)}]{RP2011}%
  \BibitemOpen
  \bibfield  {author} {\bibinfo {author} {\bibfnamefont {E.~G.}\ \bibnamefont
  {Rieffel}}\ and\ \bibinfo {author} {\bibfnamefont {W.~H.}\ \bibnamefont
  {Polak}},\ }\href@noop {} {\emph {\bibinfo {title} {Quantum Computing: A
  Gentle Introduction}}}\ (\bibinfo  {publisher} {MIT Press},\ \bibinfo {year}
  {2011})\BibitemShut {NoStop}%
\bibitem [{\citenamefont {Farhi}\ and\ \citenamefont {Gutmann}(1998)}]{FG1998}%
  \BibitemOpen
  \bibfield  {author} {\bibinfo {author} {\bibfnamefont {E.}~\bibnamefont
  {Farhi}}\ and\ \bibinfo {author} {\bibfnamefont {S.}~\bibnamefont
  {Gutmann}},\ }\bibfield  {title} {\enquote {\bibinfo {title} {Analog analogue
  of a digital quantum computation},}\ }\href {\doibase
  10.1103/PhysRevA.57.2403} {\bibfield  {journal} {\bibinfo  {journal} {Phys.
  Rev. A}\ }\textbf {\bibinfo {volume} {57}},\ \bibinfo {pages} {2403--2406}
  (\bibinfo {year} {1998})}\BibitemShut {NoStop}%
\bibitem [{\citenamefont {Shenvi}\ \emph {et~al.}(2003)\citenamefont {Shenvi},
  \citenamefont {Kempe},\ and\ \citenamefont {Whaley}}]{SKW2003}%
  \BibitemOpen
  \bibfield  {author} {\bibinfo {author} {\bibfnamefont {N.}~\bibnamefont
  {Shenvi}}, \bibinfo {author} {\bibfnamefont {J.}~\bibnamefont {Kempe}}, \
  and\ \bibinfo {author} {\bibfnamefont {K.~B.}\ \bibnamefont {Whaley}},\
  }\bibfield  {title} {\enquote {\bibinfo {title} {Quantum random-walk search
  algorithm},}\ }\href {\doibase 10.1103/PhysRevA.67.052307} {\bibfield
  {journal} {\bibinfo  {journal} {Phys. Rev. A}\ }\textbf {\bibinfo {volume}
  {67}},\ \bibinfo {pages} {052307} (\bibinfo {year} {2003})}\BibitemShut
  {NoStop}%
\bibitem [{\citenamefont {Childs}\ and\ \citenamefont
  {Goldstone}(2004)}]{CG2004}%
  \BibitemOpen
  \bibfield  {author} {\bibinfo {author} {\bibfnamefont {A.~M.}\ \bibnamefont
  {Childs}}\ and\ \bibinfo {author} {\bibfnamefont {J.}~\bibnamefont
  {Goldstone}},\ }\bibfield  {title} {\enquote {\bibinfo {title} {Spatial
  search by quantum walk},}\ }\href {\doibase 10.1103/PhysRevA.70.022314}
  {\bibfield  {journal} {\bibinfo  {journal} {Phys. Rev. A}\ }\textbf {\bibinfo
  {volume} {70}},\ \bibinfo {pages} {022314} (\bibinfo {year}
  {2004})}\BibitemShut {NoStop}%
\bibitem [{\citenamefont {Farhi}\ \emph {et~al.}(2000)\citenamefont {Farhi},
  \citenamefont {Goldstone}, \citenamefont {Gutmann},\ and\ \citenamefont
  {Sipser}}]{FGGS2000}%
  \BibitemOpen
  \bibfield  {author} {\bibinfo {author} {\bibfnamefont {E.}~\bibnamefont
  {Farhi}}, \bibinfo {author} {\bibfnamefont {J.}~\bibnamefont {Goldstone}},
  \bibinfo {author} {\bibfnamefont {S.}~\bibnamefont {Gutmann}}, \ and\
  \bibinfo {author} {\bibfnamefont {M.}~\bibnamefont {Sipser}},\ }\bibfield
  {title} {\enquote {\bibinfo {title} {Quantum computation by adiabatic
  evolution},}\ }\href@noop {} {\bibfield  {journal} {\bibinfo  {journal}
  {{a}rXiv:quant-ph/0001106}\ } (\bibinfo {year} {2000})}\BibitemShut {NoStop}%
\bibitem [{\citenamefont {Roland}\ and\ \citenamefont {Cerf}(2002)}]{RC2002}%
  \BibitemOpen
  \bibfield  {author} {\bibinfo {author} {\bibfnamefont {J.}~\bibnamefont
  {Roland}}\ and\ \bibinfo {author} {\bibfnamefont {N.~J.}\ \bibnamefont
  {Cerf}},\ }\bibfield  {title} {\enquote {\bibinfo {title} {Quantum search by
  local adiabatic evolution},}\ }\href {\doibase 10.1103/PhysRevA.65.042308}
  {\bibfield  {journal} {\bibinfo  {journal} {Phys. Rev. A}\ }\textbf {\bibinfo
  {volume} {65}},\ \bibinfo {pages} {042308} (\bibinfo {year}
  {2002})}\BibitemShut {NoStop}%
\bibitem [{\citenamefont {Roland}\ and\ \citenamefont {Cerf}(2003)}]{RC2003}%
  \BibitemOpen
  \bibfield  {author} {\bibinfo {author} {\bibfnamefont {J.}~\bibnamefont
  {Roland}}\ and\ \bibinfo {author} {\bibfnamefont {N.~J.}\ \bibnamefont
  {Cerf}},\ }\bibfield  {title} {\enquote {\bibinfo {title} {Quantum-circuit
  model of {H}amiltonian search algorithms},}\ }\href {\doibase
  10.1103/PhysRevA.68.062311} {\bibfield  {journal} {\bibinfo  {journal} {Phys.
  Rev. A}\ }\textbf {\bibinfo {volume} {68}},\ \bibinfo {pages} {062311}
  (\bibinfo {year} {2003})}\BibitemShut {NoStop}%
\bibitem [{\citenamefont {Krovi}\ \emph {et~al.}(2010)\citenamefont {Krovi},
  \citenamefont {Ozols},\ and\ \citenamefont {Roland}}]{KOR2010}%
  \BibitemOpen
  \bibfield  {author} {\bibinfo {author} {\bibfnamefont {H.}~\bibnamefont
  {Krovi}}, \bibinfo {author} {\bibfnamefont {M.}~\bibnamefont {Ozols}}, \ and\
  \bibinfo {author} {\bibfnamefont {J.}~\bibnamefont {Roland}},\ }\bibfield
  {title} {\enquote {\bibinfo {title} {Adiabatic condition and the quantum
  hitting time of markov chains},}\ }\href {\doibase
  10.1103/PhysRevA.82.022333} {\bibfield  {journal} {\bibinfo  {journal} {Phys.
  Rev. A}\ }\textbf {\bibinfo {volume} {82}},\ \bibinfo {pages} {022333}
  (\bibinfo {year} {2010})}\BibitemShut {NoStop}%
\bibitem [{\citenamefont {Meyer}\ and\ \citenamefont
  {Wong}(2015{\natexlab{a}})}]{Wong7}%
  \BibitemOpen
  \bibfield  {author} {\bibinfo {author} {\bibfnamefont {D.~A.}\ \bibnamefont
  {Meyer}}\ and\ \bibinfo {author} {\bibfnamefont {T.~G.}\ \bibnamefont
  {Wong}},\ }\bibfield  {title} {\enquote {\bibinfo {title} {Connectivity is a
  poor indicator of fast quantum search},}\ }\href {\doibase
  10.1103/PhysRevLett.114.110503} {\bibfield  {journal} {\bibinfo  {journal}
  {Phys. Rev. Lett.}\ }\textbf {\bibinfo {volume} {114}},\ \bibinfo {pages}
  {110503} (\bibinfo {year} {2015}{\natexlab{a}})}\BibitemShut {NoStop}%
\bibitem [{\citenamefont {Wong}(2016)}]{Wong9}%
  \BibitemOpen
  \bibfield  {author} {\bibinfo {author} {\bibfnamefont {T.~G.}\ \bibnamefont
  {Wong}},\ }\bibfield  {title} {\enquote {\bibinfo {title} {Spatial search by
  continuous-time quantum walk with multiple marked vertices},}\ }\href
  {\doibase 10.1007/s11128-015-1239-y} {\bibfield  {journal} {\bibinfo
  {journal} {Quantum Inf. Process.}\ ,\ \bibinfo {pages} {1--33}} (\bibinfo
  {year} {2016})},\ \bibinfo {note} {{a}rXiv:1501.07071 [quant-ph]}\BibitemShut
  {NoStop}%
\bibitem [{\citenamefont {Wong}(2015{\natexlab{a}})}]{Wong16}%
  \BibitemOpen
  \bibfield  {author} {\bibinfo {author} {\bibfnamefont {T.~G.}\ \bibnamefont
  {Wong}},\ }\bibfield  {title} {\enquote {\bibinfo {title} {Faster quantum
  walk search on a weighted graph},}\ }\href {\doibase
  10.1103/PhysRevA.92.032320} {\bibfield  {journal} {\bibinfo  {journal} {Phys.
  Rev. A}\ }\textbf {\bibinfo {volume} {92}},\ \bibinfo {pages} {032320}
  (\bibinfo {year} {2015}{\natexlab{a}})}\BibitemShut {NoStop}%
\bibitem [{\citenamefont {Meyer}\ and\ \citenamefont {Wong}(2013)}]{Wong3}%
  \BibitemOpen
  \bibfield  {author} {\bibinfo {author} {\bibfnamefont {D.~A.}\ \bibnamefont
  {Meyer}}\ and\ \bibinfo {author} {\bibfnamefont {T.~G.}\ \bibnamefont
  {Wong}},\ }\bibfield  {title} {\enquote {\bibinfo {title} {Nonlinear quantum
  search using the {G}ross-{P}itaevskii equation},}\ }\href
  {http://stacks.iop.org/1367-2630/15/i=6/a=063014} {\bibfield  {journal}
  {\bibinfo  {journal} {New J. Phys.}\ }\textbf {\bibinfo {volume} {15}},\
  \bibinfo {pages} {063014} (\bibinfo {year} {2013})}\BibitemShut {NoStop}%
\bibitem [{\citenamefont {Meyer}\ and\ \citenamefont {Wong}(2014)}]{Wong4}%
  \BibitemOpen
  \bibfield  {author} {\bibinfo {author} {\bibfnamefont {D.~A.}\ \bibnamefont
  {Meyer}}\ and\ \bibinfo {author} {\bibfnamefont {T.~G.}\ \bibnamefont
  {Wong}},\ }\bibfield  {title} {\enquote {\bibinfo {title} {Quantum search
  with general nonlinearities},}\ }\href
  {http://link.aps.org/doi/10.1103/PhysRevA.89.012312} {\bibfield  {journal}
  {\bibinfo  {journal} {Phys. Rev. A}\ }\textbf {\bibinfo {volume} {89}},\
  \bibinfo {pages} {012312} (\bibinfo {year} {2014})}\BibitemShut {NoStop}%
\bibitem [{\citenamefont {Meyer}\ and\ \citenamefont
  {Wong}(2015{\natexlab{b}})}]{Wong12}%
  \BibitemOpen
  \bibfield  {author} {\bibinfo {author} {\bibfnamefont {D.~A.}\ \bibnamefont
  {Meyer}}\ and\ \bibinfo {author} {\bibfnamefont {T.~G.}\ \bibnamefont
  {Wong}},\ }\bibfield  {title} {\enquote {\bibinfo {title} {Completeness is
  unnecessary for fast nonlinear quantum search},}\ }\href@noop {} {\bibfield
  {journal} {\bibinfo  {journal} {{a}rXiv:1502.06281 [quant-ph]}\ } (\bibinfo
  {year} {2015}{\natexlab{b}})}\BibitemShut {NoStop}%
\bibitem [{\citenamefont {Fenner}(2000)}]{Fenner2000}%
  \BibitemOpen
  \bibfield  {author} {\bibinfo {author} {\bibfnamefont {S.~A.}\ \bibnamefont
  {Fenner}},\ }\bibfield  {title} {\enquote {\bibinfo {title} {An intuitive
  {H}amiltonian for quantum search},}\ }\href@noop {} {\bibfield  {journal}
  {\bibinfo  {journal} {{a}rXiv:quant-ph/0004091}\ } (\bibinfo {year}
  {2000})}\BibitemShut {NoStop}%
\bibitem [{\citenamefont {Zimbor\'as}\ \emph {et~al.}(2013)\citenamefont
  {Zimbor\'as}, \citenamefont {Faccin}, \citenamefont {K\'ad\'ar},
  \citenamefont {Whitfield}, \citenamefont {Lanyon},\ and\ \citenamefont
  {Biamonte}}]{ZFKWLB2013}%
  \BibitemOpen
  \bibfield  {author} {\bibinfo {author} {\bibfnamefont {Z.}~\bibnamefont
  {Zimbor\'as}}, \bibinfo {author} {\bibfnamefont {M.}~\bibnamefont {Faccin}},
  \bibinfo {author} {\bibfnamefont {Z.}~\bibnamefont {K\'ad\'ar}}, \bibinfo
  {author} {\bibfnamefont {J.~D.}\ \bibnamefont {Whitfield}}, \bibinfo {author}
  {\bibfnamefont {B.~P.}\ \bibnamefont {Lanyon}}, \ and\ \bibinfo {author}
  {\bibfnamefont {J.}~\bibnamefont {Biamonte}},\ }\bibfield  {title} {\enquote
  {\bibinfo {title} {Quantum transport enhancement by time-reversal symmetry
  breaking},}\ }\href {\doibase 10.1038/srep02361} {\bibfield  {journal}
  {\bibinfo  {journal} {Sci. Rep.}\ }\textbf {\bibinfo {volume} {3}},\ \bibinfo
  {pages} {2361} (\bibinfo {year} {2013})}\BibitemShut {NoStop}%
\bibitem [{\citenamefont {Lu}\ \emph {et~al.}(2014)\citenamefont {Lu},
  \citenamefont {Biamonte}, \citenamefont {Li}, \citenamefont {Li},
  \citenamefont {Johnson}, \citenamefont {Bergholm}, \citenamefont {Faccin},
  \citenamefont {Zimbor\'as}, \citenamefont {Laflamme}, \citenamefont {Baugh},\
  and\ \citenamefont {Lloyd}}]{LBLLJBFZLBL2014}%
  \BibitemOpen
  \bibfield  {author} {\bibinfo {author} {\bibfnamefont {D.~W.}\ \bibnamefont
  {Lu}}, \bibinfo {author} {\bibfnamefont {J.~D.}\ \bibnamefont {Biamonte}},
  \bibinfo {author} {\bibfnamefont {J.}~\bibnamefont {Li}}, \bibinfo {author}
  {\bibfnamefont {H.}~\bibnamefont {Li}}, \bibinfo {author} {\bibfnamefont
  {T.~H.}\ \bibnamefont {Johnson}}, \bibinfo {author} {\bibfnamefont
  {V.}~\bibnamefont {Bergholm}}, \bibinfo {author} {\bibfnamefont
  {M.}~\bibnamefont {Faccin}}, \bibinfo {author} {\bibfnamefont
  {Z.}~\bibnamefont {Zimbor\'as}}, \bibinfo {author} {\bibfnamefont
  {R.}~\bibnamefont {Laflamme}}, \bibinfo {author} {\bibfnamefont
  {J.}~\bibnamefont {Baugh}}, \ and\ \bibinfo {author} {\bibfnamefont
  {S.}~\bibnamefont {Lloyd}},\ }\bibfield  {title} {\enquote {\bibinfo {title}
  {Chiral quantum walks},}\ }\href@noop {} {\bibfield  {journal} {\bibinfo
  {journal} {{a}rXiv:1405.6209 [quant-ph]}\ } (\bibinfo {year}
  {2014})}\BibitemShut {NoStop}%
\bibitem [{\citenamefont {Aharonov}(1999)}]{Aharonov1999}%
  \BibitemOpen
  \bibfield  {author} {\bibinfo {author} {\bibfnamefont {D.}~\bibnamefont
  {Aharonov}},\ }\enquote {\bibinfo {title} {Quantum computation},}\ in\ \href
  {\doibase 10.1142/9789812815569_0007} {\emph {\bibinfo {booktitle} {Annual
  Reviews of Computational Physics VI}}}\ (\bibinfo  {publisher} {World
  Scientific},\ \bibinfo {address} {Singapore},\ \bibinfo {year} {1999})\
  Chap.~\bibinfo {chapter} {7}, pp.\ \bibinfo {pages} {259--346}\BibitemShut
  {NoStop}%
\bibitem [{\citenamefont {Wong}\ \emph {et~al.}(2015)\citenamefont {Wong},
  \citenamefont {Tarrataca},\ and\ \citenamefont {N.}}]{Wong19}%
  \BibitemOpen
  \bibfield  {author} {\bibinfo {author} {\bibfnamefont {T.~G.}\ \bibnamefont
  {Wong}}, \bibinfo {author} {\bibfnamefont {L.}~\bibnamefont {Tarrataca}}, \
  and\ \bibinfo {author} {\bibfnamefont {Nahimov}\ \bibnamefont {N.}},\
  }\bibfield  {title} {\enquote {\bibinfo {title} {Laplacian versus adjacency
  matrix in quantum walk search},}\ }\href@noop {} {\bibfield  {journal}
  {\bibinfo  {journal} {{a}rXiv:1512.05554 [quant-ph]}\ } (\bibinfo {year}
  {2015})}\BibitemShut {NoStop}%
\bibitem [{\citenamefont {Mochon}(2007)}]{Mochon2007}%
  \BibitemOpen
  \bibfield  {author} {\bibinfo {author} {\bibfnamefont {C.}~\bibnamefont
  {Mochon}},\ }\bibfield  {title} {\enquote {\bibinfo {title} {Hamiltonian
  oracles},}\ }\href {\doibase 10.1103/PhysRevA.75.042313} {\bibfield
  {journal} {\bibinfo  {journal} {Phys. Rev. A}\ }\textbf {\bibinfo {volume}
  {75}},\ \bibinfo {pages} {042313} (\bibinfo {year} {2007})}\BibitemShut
  {NoStop}%
\bibitem [{Note1()}]{Note1}%
  \BibitemOpen
  \bibinfo {note} {This is actually the negative of Farhi and Gutmann's
  Hamiltonian, which contributes an unobservable, global phase, so we ignore it
  as did \cite {CG2004}.}\BibitemShut {Stop}%
\bibitem [{\citenamefont {Wong}(2014)}]{Wong6}%
  \BibitemOpen
  \bibfield  {author} {\bibinfo {author} {\bibfnamefont {T.~G.}\ \bibnamefont
  {Wong}},\ }\emph {\bibinfo {title} {Nonlinear Quantum Search}},\ \href
  {http://search.proquest.com/docview/1564037987?accountid=27169} {\bibinfo
  {type} {{PhD} dissertation}} (\bibinfo {year} {2014})\BibitemShut {NoStop}%
\bibitem [{\citenamefont {Messiah}(1999)}]{Messiah1999}%
  \BibitemOpen
  \bibfield  {author} {\bibinfo {author} {\bibfnamefont {A.}~\bibnamefont
  {Messiah}},\ }\href@noop {} {\emph {\bibinfo {title} {Quantum Mechanics}}}\
  (\bibinfo  {publisher} {Dover Publications},\ \bibinfo {address} {New York},\
  \bibinfo {year} {1999})\BibitemShut {NoStop}%
\bibitem [{\citenamefont {Rezakhani}\ \emph {et~al.}(2010)\citenamefont
  {Rezakhani}, \citenamefont {Pimachev},\ and\ \citenamefont
  {Lidar}}]{RPL2010}%
  \BibitemOpen
  \bibfield  {author} {\bibinfo {author} {\bibfnamefont {A.~T.}\ \bibnamefont
  {Rezakhani}}, \bibinfo {author} {\bibfnamefont {A.~K.}\ \bibnamefont
  {Pimachev}}, \ and\ \bibinfo {author} {\bibfnamefont {D.~A.}\ \bibnamefont
  {Lidar}},\ }\bibfield  {title} {\enquote {\bibinfo {title} {Accuracy versus
  run time in an adiabatic quantum search},}\ }\href {\doibase
  10.1103/PhysRevA.82.052305} {\bibfield  {journal} {\bibinfo  {journal} {Phys.
  Rev. A}\ }\textbf {\bibinfo {volume} {82}},\ \bibinfo {pages} {052305}
  (\bibinfo {year} {2010})}\BibitemShut {NoStop}%
\bibitem [{\citenamefont {Farhi}\ \emph {et~al.}(2002)\citenamefont {Farhi},
  \citenamefont {Goldstone},\ and\ \citenamefont {Gutmann}}]{FGG2002}%
  \BibitemOpen
  \bibfield  {author} {\bibinfo {author} {\bibfnamefont {E.}~\bibnamefont
  {Farhi}}, \bibinfo {author} {\bibfnamefont {J.}~\bibnamefont {Goldstone}}, \
  and\ \bibinfo {author} {\bibfnamefont {S.}~\bibnamefont {Gutmann}},\
  }\bibfield  {title} {\enquote {\bibinfo {title} {Quantum adiabatic evolution
  algorithms with different paths},}\ }\href@noop {} {\bibfield  {journal}
  {\bibinfo  {journal} {{a}rXiv:quant-ph/0208135}\ } (\bibinfo {year}
  {2002})}\BibitemShut {NoStop}%
\bibitem [{\citenamefont {Hofmann}\ and\ \citenamefont
  {Schaller}(2014)}]{HS2014}%
  \BibitemOpen
  \bibfield  {author} {\bibinfo {author} {\bibfnamefont {M.}~\bibnamefont
  {Hofmann}}\ and\ \bibinfo {author} {\bibfnamefont {G.}~\bibnamefont
  {Schaller}},\ }\bibfield  {title} {\enquote {\bibinfo {title} {Probing
  nonlinear adiabatic paths with a universal integrator},}\ }\href {\doibase
  10.1103/PhysRevA.89.032308} {\bibfield  {journal} {\bibinfo  {journal} {Phys.
  Rev. A}\ }\textbf {\bibinfo {volume} {89}},\ \bibinfo {pages} {032308}
  (\bibinfo {year} {2014})}\BibitemShut {NoStop}%
\bibitem [{\citenamefont {Crosson}\ \emph {et~al.}(2014)\citenamefont
  {Crosson}, \citenamefont {Farhi}, \citenamefont {Lin}, \citenamefont {Lin},\
  and\ \citenamefont {Shor}}]{Crosson2014}%
  \BibitemOpen
  \bibfield  {author} {\bibinfo {author} {\bibfnamefont {E.}~\bibnamefont
  {Crosson}}, \bibinfo {author} {\bibfnamefont {E.}~\bibnamefont {Farhi}},
  \bibinfo {author} {\bibfnamefont {C.~Y.-Y.}\ \bibnamefont {Lin}}, \bibinfo
  {author} {\bibfnamefont {H.-H.}\ \bibnamefont {Lin}}, \ and\ \bibinfo
  {author} {\bibfnamefont {P.}~\bibnamefont {Shor}},\ }\bibfield  {title}
  {\enquote {\bibinfo {title} {Different strategies for optimization using the
  quantum adiabatic algorithm},}\ }\href@noop {} {\bibfield  {journal}
  {\bibinfo  {journal} {{a}rXiv:1401.7320 [quant-ph]}\ } (\bibinfo {year}
  {2014})}\BibitemShut {NoStop}%
\bibitem [{\citenamefont {Bennett}\ \emph {et~al.}(1997)\citenamefont
  {Bennett}, \citenamefont {Bernstein}, \citenamefont {Brassard},\ and\
  \citenamefont {Vazirani}}]{BBBV1997}%
  \BibitemOpen
  \bibfield  {author} {\bibinfo {author} {\bibfnamefont {C.~H.}\ \bibnamefont
  {Bennett}}, \bibinfo {author} {\bibfnamefont {E.}~\bibnamefont {Bernstein}},
  \bibinfo {author} {\bibfnamefont {G.}~\bibnamefont {Brassard}}, \ and\
  \bibinfo {author} {\bibfnamefont {U.}~\bibnamefont {Vazirani}},\ }\bibfield
  {title} {\enquote {\bibinfo {title} {Strengths and weaknesses of quantum
  computing},}\ }\href {\doibase 10.1137/S0097539796300933} {\bibfield
  {journal} {\bibinfo  {journal} {SIAM J. Comput.}\ }\textbf {\bibinfo {volume}
  {26}},\ \bibinfo {pages} {1510--1523} (\bibinfo {year} {1997})}\BibitemShut
  {NoStop}%
\bibitem [{\citenamefont {Aharonov}\ \emph {et~al.}(2004)\citenamefont
  {Aharonov}, \citenamefont {van Dam}, \citenamefont {Kempe}, \citenamefont
  {Landau}, \citenamefont {Lloyd},\ and\ \citenamefont {Regev}}]{Aharonov2004}%
  \BibitemOpen
  \bibfield  {author} {\bibinfo {author} {\bibfnamefont {D.}~\bibnamefont
  {Aharonov}}, \bibinfo {author} {\bibfnamefont {W.}~\bibnamefont {van Dam}},
  \bibinfo {author} {\bibfnamefont {J.}~\bibnamefont {Kempe}}, \bibinfo
  {author} {\bibfnamefont {Z.}~\bibnamefont {Landau}}, \bibinfo {author}
  {\bibfnamefont {S.}~\bibnamefont {Lloyd}}, \ and\ \bibinfo {author}
  {\bibfnamefont {O.}~\bibnamefont {Regev}},\ }\bibfield  {title} {\enquote
  {\bibinfo {title} {Adiabatic quantum computation is equivalent to standard
  quantum computation},}\ }in\ \href {\doibase 10.1109/FOCS.2004.8} {\emph
  {\bibinfo {booktitle} {Proceedings of the 45th Annual IEEE Symposium on
  Foundations of Computer Science}}},\ \bibinfo {series and number} {FOCS '04}\
  (\bibinfo  {publisher} {IEEE Computer Society},\ \bibinfo {year} {2004})\
  pp.\ \bibinfo {pages} {42--51}\BibitemShut {NoStop}%
\bibitem [{\citenamefont {Childs}\ \emph {et~al.}(2013)\citenamefont {Childs},
  \citenamefont {Gosset},\ and\ \citenamefont {Webb}}]{Childs2013}%
  \BibitemOpen
  \bibfield  {author} {\bibinfo {author} {\bibfnamefont {A.~M.}\ \bibnamefont
  {Childs}}, \bibinfo {author} {\bibfnamefont {D.}~\bibnamefont {Gosset}}, \
  and\ \bibinfo {author} {\bibfnamefont {Z.}~\bibnamefont {Webb}},\ }\bibfield
  {title} {\enquote {\bibinfo {title} {Universal computation by multiparticle
  quantum walk},}\ }\href {\doibase 10.1126/science.1229957} {\bibfield
  {journal} {\bibinfo  {journal} {Science}\ }\textbf {\bibinfo {volume}
  {339}},\ \bibinfo {pages} {791--794} (\bibinfo {year} {2013})}\BibitemShut
  {NoStop}%
\bibitem [{Note2()}]{Note2}%
  \BibitemOpen
  \bibinfo {note} {Fenner's Hamiltonian in \cite {Fenner2000} is actually twice
  this; ours is closer in form and runtime to Farhi and Gutmann's.}\BibitemShut
  {Stop}%
\bibitem [{\citenamefont {Wong}(2015{\natexlab{b}})}]{Wong14}%
  \BibitemOpen
  \bibfield  {author} {\bibinfo {author} {\bibfnamefont {T.~G.}\ \bibnamefont
  {Wong}},\ }\bibfield  {title} {\enquote {\bibinfo {title} {Quantum walk
  search with time-reversal symmetry breaking},}\ }\href {\doibase
  10.1088/1751-8113/48/40/405303} {\bibfield  {journal} {\bibinfo  {journal}
  {J. Phys. A: Math. Theor.}\ }\textbf {\bibinfo {volume} {48}},\ \bibinfo
  {pages} {405303} (\bibinfo {year} {2015}{\natexlab{b}})}\BibitemShut
  {NoStop}%
\end{thebibliography}%

\end{document}